\pdfoutput=1
\documentclass[prl,floatfix,twocolumn,showkeys]{revtex4-1}

\usepackage{color} 
\usepackage{graphicx} 
\usepackage{amsmath,amssymb} 

\begin{document}

\author{Diego Prada-Gracia $^1$}
\author{Roman Shevchuk $^1$}

\author{Peter Hamm $^2$}
\email{phamm@pci.uzh.ch}

\author{Francesco Rao $^1$}
\email{francesco.rao@frias.uni-freiburg.de}

\affiliation{$^1$ Freiburg Institute for Advanced Studies, School of Soft
Matter Research, Freiburg im Breisgau, Germany.}
\affiliation{$^2$ Institute of Physical Chemistry, University of Zurich, Zurich, Switzerland.}

%\title{Water structural inhomogeneities in temperature reveal a funneled energy
%landscape}

\title{Towards a microscopic description of the free-energy landscape of water}

\begin{abstract} 

  Free-energy landscape theory is often used to describe complex molecular
  systems.  Here, a microscopic description of water structure and dynamics based
  on configuration-space-networks and molecular dynamics simulations of the
  TIP4P/2005 model is applied to investigate the free-energy landscape of
  water.  The latter is built on top of a large set of water microstates
  describing the kinetic stability of local hydrogen-bond arrangements up to
  the second solvation shell. In temperature space, the landscape displays
  three regions with an overall different organization. At ambient conditions,
  the free-energy surface is characterized by structural inhomogeneities with
  multiple, structurally well defined, short-lived basins of attraction.  Below
  around ambient temperature, the liquid rapidly becomes homogeneous. In this
  regime, the landscape is funneled-like, with fully-coordinated water
  arrangements at the bottom of the funnel. Finally, a third region develops
  below the temperature of maximal compressibility (Widom line) where the
  funnel becomes steeper with few interconversions between microstates other
  than the fully coordinated ones.  Our results present a viable a way to
  manage the complexity of water structure and dynamics, connecting microscopic
  properties to its ensemble behavior.

%Here, we study the behavior of water in temperature in terms of structural
%inhomogeneities.  Such inhomogeneities are short-lived local water
%configurations of the size of at least two-solvation shells, the study of which
%have recently attracted both experimentalists and theoreticians. The analysis
%reveals a free energy landscape funnel like with two different regimes clearly
%distinguished by the kinetics and thermodynamics of the second solvation shell:
%a low temperature regime where a predominant class of microstates organized as
%a single basin of attraction enthalpically stabilized, and a high temperature
%regime where the second solvation shell pass through different structural
%meta-stable states entropically stabilized.
%
%\red{(not with parameters as Q, crystallization, etch) This behavior emerges
%from the enthalpy-entropy balance.}

\end{abstract}

\date{\today}

\maketitle

%\blue{--- use past tense --- run spell check --- comment on the Widom line}

\section{Introduction}

Water is the most studied liquid in nature. Notwithstanding the effort, the
debate is still open when it comes to the connection between atomic properties
and its ensemble behavior. As an example, the presence of a density maximum at
a temperature of 4 degrees Celsius \cite{Waller1684Essayes}
%\cite{Rontgen1892Ueber,Mishima1998Relationship}, 
is qualitatively interpreted as the result of two competing driving forces,
i.e., the directionality of hydrogen bonds, favoring tetrahedral
(lower-density) water arrangements which are enthalpically stabilized, and
entropy maximization by non-directional interactions and disorder, resulting in
a closer packing (higher-density).  Recent experimental and computational
evidences pointed out that at ambient conditions these arrangements have
%distinct structural and dynamical properties \cite{Rezus2007Observation,
distinct structural and dynamical properties \cite{Huang2009Inhomogeneous,
Rao2010Structural, GarrettRoe2011ThreeDimensional}.  These findings might be
related to a century-old idea describing water anomalies as an emergent
property of a mixture-like liquid \cite{Rontgen1892Ueber,
Errington2001Relationship, Cuthbertson2011Mixturelike, Wikfeldt2011Enhanced}.
This picture is far from conclusive, with other groups providing counter
evidences to these ideas
\cite{Clark2010Smallangle,Moore2011Structural,Limmer2011Putative}. 

In other fields, a useful approach to clarify the connection between
microscopic properties and ensemble behavior makes use of the free-energy
paradigm.  Energy landscape theory have demonstrated to be a successful
approach for the study of the structure and dynamics of complex molecular
systems
\cite{Goldstein1969Viscous,Stillinger1995Topographic,Wolynes1995Navigating,Onuchic1997Theory,
Wales1998Archetypal}.  Within this framework, molecular dynamics is interpreted
as a trajectory on the multidimensional free-energy surface. The latter is made
up of many valleys connected by saddles, suggesting that system dynamics can be
divided into intravalley and intervalley motions \cite{Frauenfelder1991Energy}.
The former represent the oscillations around local minima, while the latter
involve barrier crossings from one minimum to another \cite{Rao2010Protein}.
Most descriptions of the energy surface of water have been limited to ensemble
properties or clusters of a few tens of molecules
\cite{Angell1995Formation,Wales1998Global,Ludwig2001Water}. Recently, a mapping
of the free-energy surface of bulk water from a more microscopic perspective,
an extension of configuration-space-networks \cite{Rao2004Protein,
Gfeller2007Complex}, was proposed \cite{Rao2010Structural}. Network
configurations (i.e. nodes) and links represented local water hydrogen-bond
arrangements with an extension of two solvation shells and the transitions
among them as sampled by molecular dynamics simulations, respectively.  This
approach represents a reductionist strategy (i.e., focusing on the microscopic
behavior) to describe the free-energy landscape of water.

In this work, network analysis is extended beyond ambient temperature by
running extensive molecular dynamics simulations of the TIP4P/2005 water model
from 340K to the supercooled regime. As the system is cooled down, the
free-energy landscape is characterized by three different regimes. At
temperatures up to the physiological one, the free-energy presents several,
short-lived, basins of attraction. Below this temperature, the liquid becomes
homogeneous with the development of a funnel-like energy landscape. Finally, by
passing the temperature of maximum compressibility the funnel rapidly becomes
more pronounced with a very strong bias towards fully coordinated structures. 

Within the context of water structure and dynamics, our results provide a
reductionist approach linking microscopic behavior to water ensemble
properties.

\section{Methods}

{\bf Molecular dynamics simulations.} All molecular dynamics simulations were
run with the program GROMACS \cite{VanDerSpoel2005GROMACS,Hess2008GROMACS}. The
water box consisted of 1024 TIP4P/2005 \cite{Abascal2005General} water molecules in the NPT
ensemble with pressure of 1 atm and temperatures ranging from 190 K to 340 K
with steps of 10 K. The integration time-step was 2 fs, saving coordinates
every 4 fs.  Equilibration was run for 15 ns, followed by a 2 ns long run to
collect the data. For temperatures below 240K the equilibration time was
elongated up to 25 ns.  During the production runs, the system was coupled to a
Berendsen barostat \cite{Berendsen1984Molecular} and a velocity rescale
thermostat \cite{Bussi2007Canonical} with  coupling times $\tau_P=1.0$ ps and
$\tau_T=1.0$ ps, respectively. Long range electrostatics was computed with PME
\cite{Darden1993Particle}, using a $1.0$ nm cutoff for all non-bonded
interactions. Calculations with the TIP3P \cite{Jorgensen1983Comparison} and TIP5P \cite{Mahoney2000Fivesite}
models were carried out using the same protocol.
 
{\bf Hydrogen bond definitions.} The hydrogen bond definition introduced by
Skinner and collaborators was used \cite{Kumar2007Hydrogen}. In this approach
the orbital occupancy is parametrized in terms of the intermolecular distance
$r$ between the acceptor O and donor H and the angle $\psi$ that the O $\cdots$
H ray makes with the out-of-plane unit vector of the acceptor molecule using
the formula:
\begin{equation} N(r,\psi)=exp(-r/0.343)(7.1-0.050\psi+0.00021\psi^2)
\end{equation}
As in the original work, two molecules are considered to be hydrogen-bonded if
the occupancy $N(r,\psi)$ exceeds a certain threshold ($N>0.0085$).  To check
the robustness of our results with the hydrogen bond definition, the analysis
was repeated considering the conventional inter-oxygen distance and angle O-H-O
(cutoffs of 3.5 \AA\ and 30 degrees, respectively~\cite{Luzar1996Effect}). 

{\bf Tetrahedral order parameter.} The tetrahedral order parameter for a water
molecule $i$ is calculated as
\begin{equation}
	q_{i}=1-\frac{3}{8}\sum_{j=1}^{3}\sum_{k=j+1}^{4}\Big(cos\psi_{jik}+\frac{1}{3}\Big)^{2},
\end{equation}
% where $j$ and $k$ are any of the four nearest water molecules of $i$
where $\psi_{jik}$ is the angle formed by their oxygens
\cite{Errington2001Relationship}. The averaged value of this order parameter
over an ensemble of water molecules is denoted as $Q$.

\section{Theory}

Configuration-space-networks (often referred as Markov-State-Models) provide
high resolution free-energy landscape representations of complex molecular
systems \cite{Rao2004Protein, Krivov2004Hidden, Gfeller2007Complex, Noe2007Hierarchical,
  Chodera2007Automatic,Noe2008Transition, PradaGracia2009Exploring,
Rao2010Protein}.  The main idea behind this approach is to map the molecular
dynamics onto a transition network where nodes and links represent system
configurations and the transitions between them, respectively. System
configurations are \emph{microstates}, i.e.  very small portions of the
accessible configuration space. Within this approach a variety of molecular
trajectories from molecular dynamics simulations
\cite{Rao2004Protein,Gfeller2007Complex,Rao2010Protein} to single molecule
experiments \cite{Baba2007Construction, Schuetz2010Free,
Baba2011Multidimensional} can be analyzed. The great advantage of
configuration-space-networks with respect to conventional methods is that
free-energy representations are obtained without the use of 
projections onto arbitrarily chosen order parameters \cite{Rao2004Protein,
Krivov2004Hidden}.

\begin{figure}
  \includegraphics[width=80.0mm]{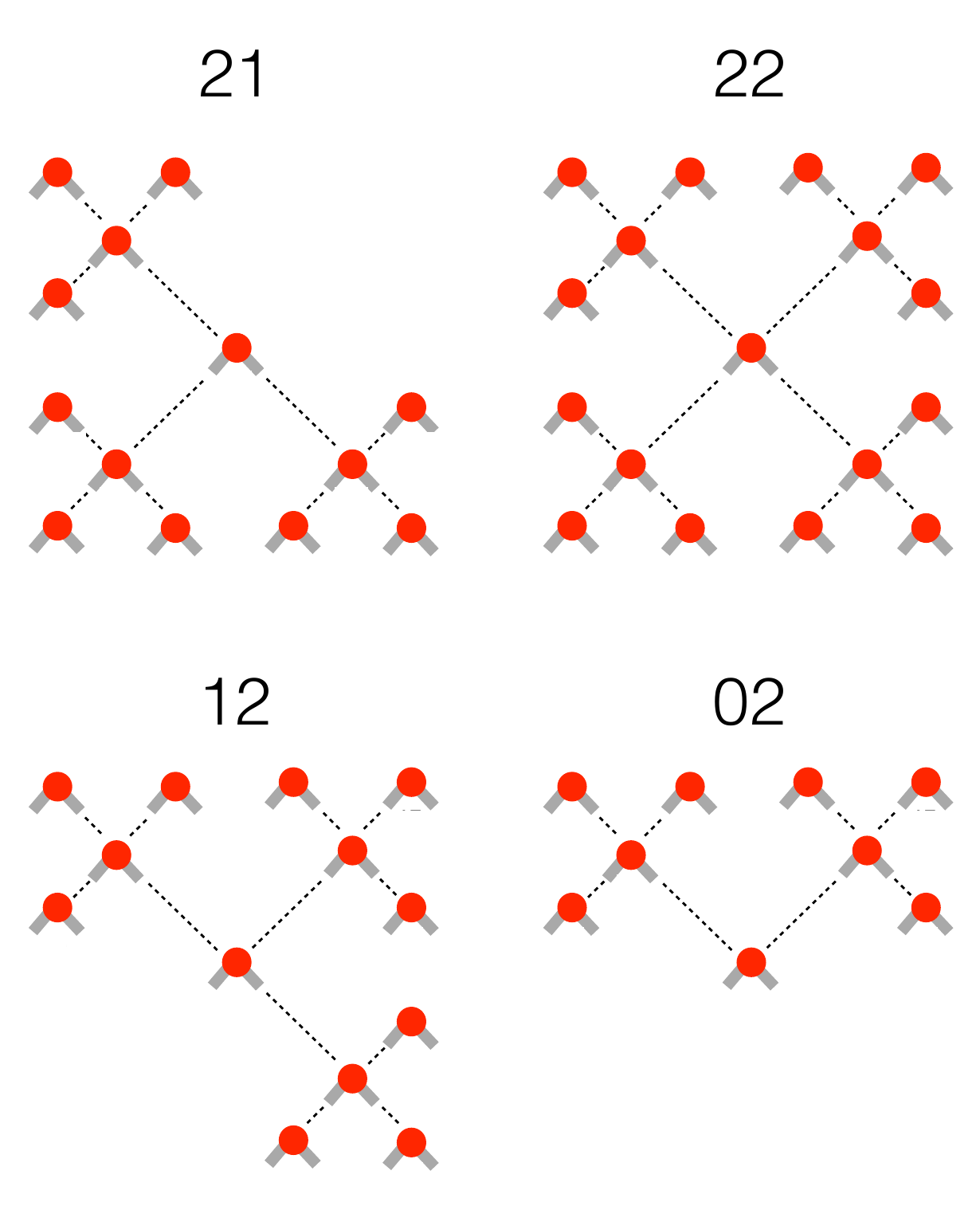}
  \caption{Representative water microstates belonging to the four most
    populated gradient-clusters at 300K.  Hydrogen bonds are
    represented as dashed lines. For simplicity of reference, each of the four
  configurations is classified by two numbers, indicating the number of donors
and acceptors of the central water molecule (e.g. 21 stands for two donors and
one acceptor).}
  \label{fig:microstates}
\end{figure}

Very recently, we showed that configuration-space-networks can be used to map
the free-energy landscape of bulk water at ambient conditions
\cite{Rao2010Structural}. In this case, microstates are defined by
hydrogen-bond connectivity patterns including the first and second solvation
shells of a given water molecule (see Fig.~\ref{fig:microstates}). 
%In this way a microstate is encoded into a text string of 17 digits: the
%central water molecule, four first neighbors and 12 second neighbors. 
The characterization of the hydrogen-bond pattern is built in a way that takes
into account of the indistinguishable nature of water molecules and all the
possible permutations (see Ref.~\cite{Rao2010Structural} for details).
Consequently, the configuration-space-network of water is built by following
the evolution of the microstates of each water molecule of the simulation box,
resulting in 1024 trajectories of length 2 ns for the present case.
Network links are direct molecular transitions between water microstates of the
same molecule  at a temporal resolution imposed by the saving frequency of
Cartesian coordinates (4~fs in the present work). By recording link
populations, the final configuration-space-network is equivalent to a
transition matrix.

As observed elsewhere \cite{Krivov2004Hidden, Gfeller2007Complex,
  Rao2010Protein}, the transition network resulting from this analysis
synthetically encodes the complex organization of the underlying
free-energy landscape.  Specifically, densely connected regions of the
network correspond to free-energy basins, i.e., metastable regions of
the configuration space.  Several algorithms can be used to extract
this information, including the max flow theorem
\cite{Krivov2004Hidden}, random walks \cite{vanDongen2000Graph,
  Gfeller2007Complex} or transition gradient analysis
\cite{PradaGracia2009Exploring, Rao2010Local}. All these approaches
aim to clusterize the network into kinetically and structurally well
defined basins of attraction.

\begin{figure}
  \includegraphics[width=50.0mm]{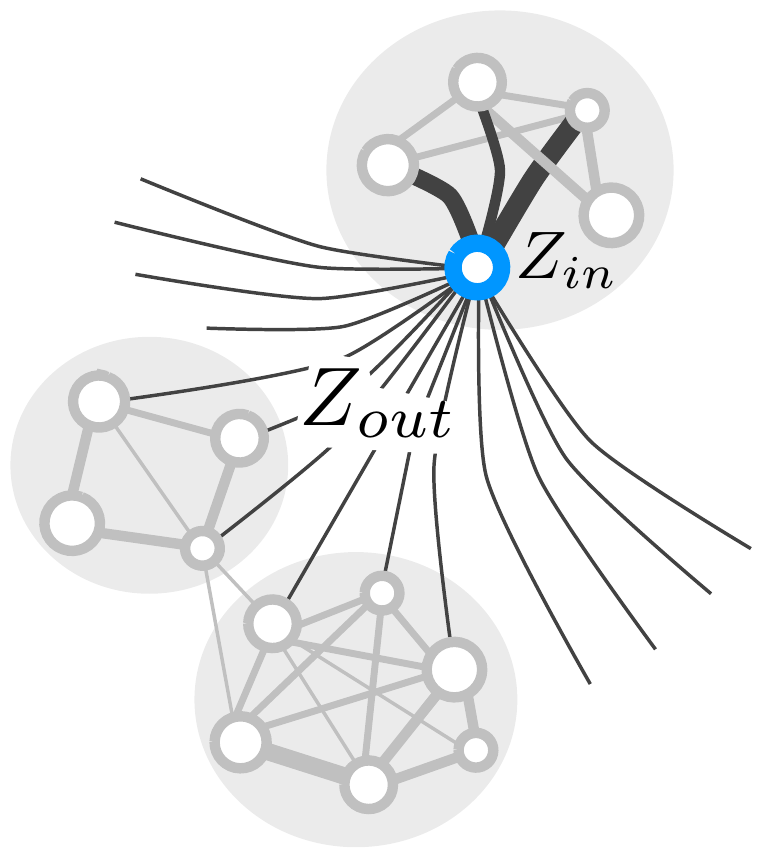}
  \caption{Configuration-space-networks. Pictorial representation of the
    relative balance between intra-basin ($Z_{in}$) and inter-basins
    ($Z_{out}$) transition probabilities from the point of view of a node (in
    blue). Gray regions represent free-energy basins of attraction as detected
    by the gradient-algorithm \cite{PradaGracia2009Exploring, Rao2010Local} (see Theory for details). }
  \label{fig:theory}
\end{figure}

In enthalpy driven free-energy landscapes, of which proteins are an archetypal
example,  the transition probability to stay inside a given basin $Z_{in}$ is
much larger then the probability to hop outside $Z_{out}$
\cite{Krivov2004Hidden,Rao2010Protein}. That is, basin hoping is a rare event.
Moreover, the number of neighboring basins is usually very limited, with the
emergence of well defined transition pathways \cite{Rao2004Protein,
Rao2010Protein,Noe2007Hierarchical}. This is not the case for water
\cite{Rao2010Structural}. Being a liquid, it is mainly characterized by
entropic basins of attraction.  As illustrated in Fig.~\ref{fig:theory},
$Z_{in}$ and $Z_{out}$ become comparable because the cumulative of the
\emph{many} small inter-basin transition probabilities ($Z_{out}$) is similar
to the \emph{few} highly populated intra-basin relaxations ($Z_{in}$). In other
words, the probability to leave the basin is similar to stay in it. This
observation would lead to the conclusion that, at the atomic level, water does
not have any type of \emph{configurational} selection. However, this is not
true when considering all the contributions to $Z_{out}$ separately:
\begin{equation} Z_{out}=\sum_i Z_{out}^{(i)} \end{equation}
%
%Disentangling individual $Z_{out}$ contributions is useful to understand water
%structure and dynamics. This approach was used in our previous work to
%investigate water structural inhomogeneities \cite{Rao2010Structural}.
%
Structural inhomogeneities, i.e., configurational selection, emerge because 
  \begin{equation} max\left(Z_{in}^{(i)}\right) \gg
    max\left(Z_{out}^{(i)}\right), \end{equation} 
meaning that the probability of an intra-basin transition is larger than hoping
to any other \emph{specific} basin. When this condition holds, the environment
of a given water molecule alternatively adopts a number of different
configurations, each of them characterized by a specific free-energy basin of
attraction. This is an emergent property of water at ambient temperature
\cite{Rao2010Structural}. 
%Hence, although the probability to stay in a basin is comparable to the one to
%hop away ($Z_{in}\approx Z_{out}$), the probabilities to hop to specific
%molecular rearrangements ($Z_{out}^{(i)}$) are much smaller than staying in
%the present basin ($Z_{in}$). 

Such structural inhomogeneities are found by using a gradient-based approach to
cluster the transition network, grouping together water microstates belonging
to the same steepest descent free-energy pathway
\cite{PradaGracia2009Exploring,Rao2010Local,Rao2010Structural}. From an
operative point of view, the algorithm works on a per-node basis by deleting
all the links (transitions) but the most visited one (which represents the
local direction of the gradient).  When applied to the whole network, the
algorithm provides a set of disconnected trees, each of them representing a
collective pathway of relaxation to the bottom of the \emph{local} free-energy
basin of attraction (gradient-cluster, gray regions in Fig.~\ref{fig:theory}).
Each gradient-cluster represents a structurally and kinetically well defined
molecular arrangement with an extension of up to two solvation shells
\cite{Rao2010Structural}. 

In this contribution, the behavior of these free-energy basins is investigated
in temperature space to elucidate the global organization of the free-energy
landscape, including the relationship between microscopic behavior and ensemble
properties.

\section{Results and Discussion}

\subsection{The free-energy landscape of water}

\begin{figure}
  \includegraphics[width=80.0mm]{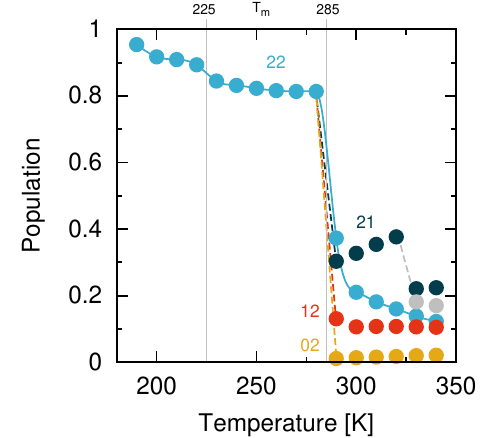}
  \caption{Population of the most visited gradient-clusters as a function of
  temperature for the TIP4P/2005 model. Vertical lines correspond to 225K and
285K. $T_m$ indicates the melting temperature of the model at around 250K
\cite{Abascal2005General}.}
  \label{fig:populations}
\end{figure}

Molecular dynamics simulations of the TIP4P/2005 water model at temperatures
from 190K to 340K were run and their corresponding configuration-space-networks
built.   For each of these networks, we looked for the free-energy basins
characterizing local water arrangements by means of a gradient-cluster analysis
(see Theory) \cite{PradaGracia2009Exploring, Rao2010Local, Rao2010Structural}.
In Fig.~\ref{fig:populations}, the population of the most visited
gradient-clusters is shown as a function of temperature.  At temperatures
larger than 285K, several free-energy basins of attraction are found in
agreement with previous analysis on the SPC model at 300K
\cite{Rao2010Structural}.  The structural configurations at the bottom of the
most visited free-energy basins are pictorially represented in
Fig.~\ref{fig:microstates}.  They correspond to the following hydrogen-bond
configurations of the central water molecule: 2 donors, 1 acceptor (21, dark
blue in Fig.~\ref{fig:populations}, population of 0.32 at 300K); 2 donors, 2
acceptors (22, light blue, 0.21); 1 donor, 2 acceptors (12, red, 0.13); 0
donors, 2 acceptors (02, yellow, 0.01). At the highest temperatures a fifth
basin appears being characterized by a 11 first solvation shell (gray). This
acceptor/donor representation is adopted for simplicity but the contribution of
the second solvation shell organization is strictly needed when it comes to
correctly characterize the free-energy basins (e.g. there are basins of
attraction with the same first shell but different second shell
\cite{Rao2010Structural}).

In this temperature range the liquid is \emph{inhomogeneous} in the sense that
the local environment of a water molecule interconverts  between configurations
with distinct structural and dynamical properties. Those represent short-lived
metastable arrangements with sub-ps lifetime \cite{Rao2010Structural,
GarrettRoe2011ThreeDimensional}.

Below 285K this property is lost as shown by the rapid increase of the
population of the 22 gradient-cluster to a value larger than 0.8 (light blue in
Fig.~\ref{fig:populations}).  As such, all highly populated gradient-clusters
collapse to 22, being the only largely populated free-energy basin. The
population of this basin is almost constant till 225K. In this regime, the
liquid is \emph{homogeneous} and the free-energy landscape resembles a funnel,
with the fully-coordinated configuration 22 at the bottom of it (see also
Section C).  The funnel behavior emerges because 22 becomes a global attractor
of the dynamics as it is the case for the native state in protein folding
\cite{dill1997levinthal}.  Still, a cumulative population of 0.2 split into
roughly six basins survives.  These configurations are rich of 4-fold
hydrogen-bond loops, slowly interconverting with the fully coordinated
configuration.

Below 225K, i.e. roughly below the temperature of maximum compressibility
(estimated to be around 230K \cite{Abascal2010Widom}), the entire landscape
collapses onto 22 with a much more pronounced funnel behavior (see also Section
C).  Interestingly, the temperature of maximum compressibility is considered by
some as the Widom line, i.e.  the propagation of a liquid-liquid critical point
located at higher pressure \cite{Mishima1998Relationship, Abascal2010Widom,
Wikfeldt2011Enhanced}. If this is so, this water regime would be connected to
the mentioned transition.  In this temperature region the density assumes its
minimum value (see Fig.~\ref{fig:topology}d). For this reason, we refer to this
temperature segment as the \emph{low-density} homogeneous regime of the liquid. 

\begin{figure}
  \includegraphics[width=70.0mm]{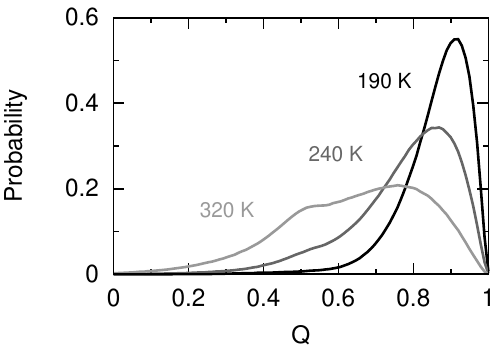}
  \caption{Distribution of the tetrahedral order parameter $Q$ for
    the three regimes.}
  \label{fig:q}
\end{figure}

It is interesting to compare these regimes with the distribution of the average
tetrahedral order parameter $Q$ (see Methods) \cite{Errington2001Relationship}.
In Fig.~\ref{fig:q} data for 320K, 240K and 190K is shown. At around ambient
conditions the distribution is bi-modal (light gray), indicating that the
liquid assumes both ordered and disordered atomic arrangements. This property
is lost at lower temperatures where the distribution becomes uni-modal (gray)
with a small population for values close to 0.5. This sub-population disappears
below the temperature of maximum compressibility, resulting in a sharply peaked
distribution. The shape shift from bi-modal to uni-modal is in good agreement
with the change from inhomogeneous regime to the homogeneous one.  

Summarizing this section, three different regimes for the liquid phase of water
were found. Each of these regions is characterized by a specific organization
of the underlying free-energy landscape as shown by the temperature dependence
of the populations of the major free-energy basins
(Fig.~\ref{fig:populations}).

\subsection{Network properties and relation to the density anomaly}

\begin{figure}
  \includegraphics[width=85.0mm]{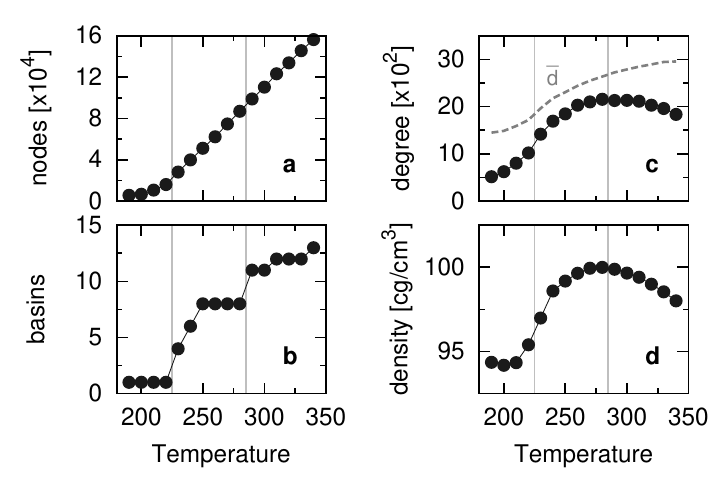} \caption{Topology of the
    configuration-space-network as a function of temperature. (a) Number of
    nodes; (b) Number of gradient-clusters with a population larger than 0.01;
    (c) Number of connections of the 22 node. For comparison, the average
    number of connections per node $\bar d$ is shown as a dashed line (in this
    case the multiplicative factor is 1 and not $\times 10^2$); (d) density.
    Vertical lines correspond to 225K and 285K.}
  \label{fig:topology}
\end{figure}

\begin{figure*}
  \includegraphics[width=120.0mm]{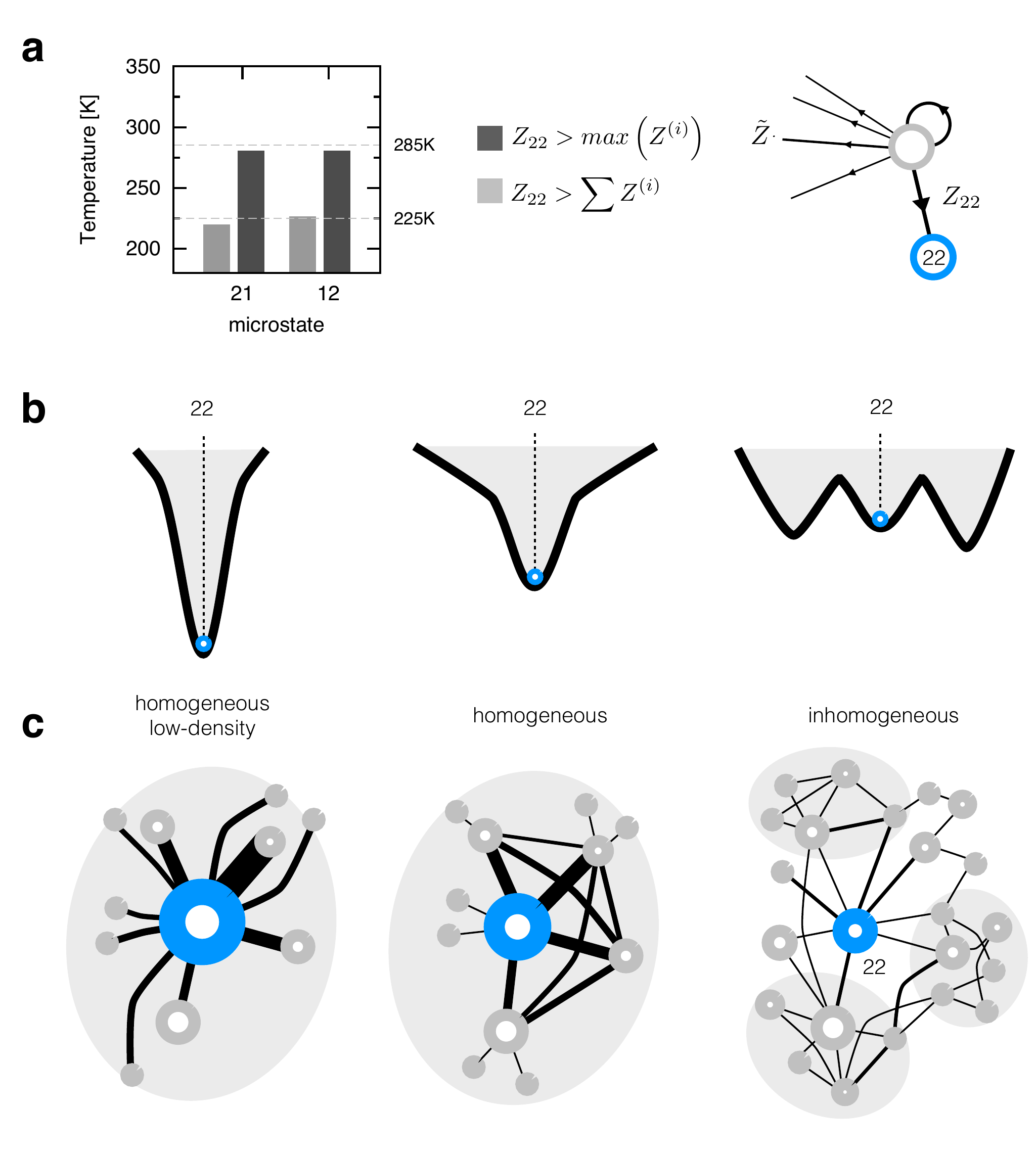}
  \caption{Schematic representations of the three regimes of liquid water; (a)
  temperatures at which there is a change in the transition probability maximum
for the 21 and 12 microstates (see Results for details); (b) free-energy landscape representation; (c)
network representation.}
  \label{fig:regimes}
\end{figure*}

As a function of temperature, the number of visited microstates (i.e. nodes) is
increasing monotonously, as shown by Fig.~\ref{fig:topology}a. That is, the
higher the temperature the larger the portion of the configuration space
visited by the molecular dynamics simulation. Above 225K the relation is
  linear but below this temperature the number of microstates changes in a
non-linear way, visiting in proportion a smaller fraction of the configuration
space. This behavior might be related to the breakdown
of the Einstein diffusion relationship below the temperature of maximum
compressibility as observed for the ST2 model \cite{Kumar2007Relation}.

In Fig.~\ref{fig:topology}b, the number of gradient-clusters with a population
larger than 0.01 as a function of temperature is shown. The data presents a
step wise behavior, correlating very well with the presence of the three
regimes. Interestingly, the number of gradient-clusters is mostly constant in
the two low-temperature regimes with only one free-energy basin below 225K.

From a network topology point of view, the number of connections per node
(degree) grows with temperature, going from an average value $\bar d$ of 14.47
to 29.55 at 190K and 340K, respectively (dashed line in
Fig.~\ref{fig:topology}c). This is not the case for the node degree of the
microstate 22. As shown in Fig.~\ref{fig:topology}c, the degree increases up to
around 285K. Then it starts to decrease where the liquid changes from the
homogeneous regime to the inhomogeneous one. 
%In both connectivity mesaures,
%there is a change in concavity at 225K in agreement with the non-linear
%behavior of the number of visited microstates (panel a).

Comparison with density (Fig.~\ref{fig:topology}d) shows a remarkable
correlation. With a Pearson coefficient of 0.98, the behavior of the node
degree of microstate 22 correlates with the density anomaly. This seems an
interesting fact connecting an ensemble property like the density to a purely
microscopic quantity, i.e.  the number of accessible transitions from the fully
coordinated configuration 22.

\subsection{The origin of a funneled energy landscape}

In this section, the transition network modifications corresponding to the
three regimes are illustrated.  In the higher temperatures regime, structural
inhomogeneities emerge because the maximum of the transition probability
$max\left(Z^{(i)}\right)$ points towards the attractor of the basin (e.g.
$\tilde Z$ in the pictorial representation of Fig.~\ref{fig:regimes}a). This is
not the case below 285K where the transition to 22 ($Z_{22}$) becomes the
maximum of the transition probability for many nodes acting as attractors at higher
temperatures.  Fig.~\ref{fig:regimes}a shows the temperature at which
$Z_{22}=max\left(Z^{(i)}\right)$  for the relevant microstates 21 and 12 (dark
gray bars).  Relaxing directly to 22, they do not build basins of attraction
anymore.  For nodes not directly connected to 22 the relaxation process to it
goes through two or more steps like for 02. Consequently, a free-energy
landscape characterized by a single predominant minimum (22) develops. This
type of landscapes recall the well-known funnel-landscape paradigm applied to
protein folding
\cite{Leopold1992Protein,Bryngelson1995Funnels,dill1997levinthal}.

Below 225K, $Z_{22}$ drives the dynamics in a even stronger way being the
corresponding transition larger than the cumulative of all other
transitions (gray bars in Fig.~\ref{fig:regimes}a). In other words, every time a
water molecule assumes a configuration different from 22, the probability to go
back to 22 is larger with respect to the cumulative of any other transition.

From a qualitative point of view the three regimes of the free-energy landscape
are represented in Fig.~\ref{fig:regimes}b (in panel c a pictorial
representation of the underlying network). 

%In the inhomogeneous regime, the landscape presents
%several, mainly entropically stabilized basins \cite{Rao2010Structural}. Below
%285K, the landscape becomes funneled towards the fully coordinated 22
%microstate. At this point, transitions to nodes other than 22 are possible with
%transient paths in the neighboring of 22 (homogeneous regime). Finally, in
%the low-density regime below 225K the funnel becomes more pronounced with a
%star-like topology of the underlying network.

%In folding, the unfolded state is represented by an ensemble of metastable
%configurations with a more entropic (larger volume of space) compared to  the
%folded state. This unfolded set of conformations has a hierarchical
%organization with a topology driving the system to the bottom of the funnel
%with a higher probability. Arranged in convergent kinetic pathways guiding the
%system to a unique, stable, 'native conformation'

\subsection{Robustness and limitations of the current methodology}

\begin{figure}
  \includegraphics[width=80.0mm]{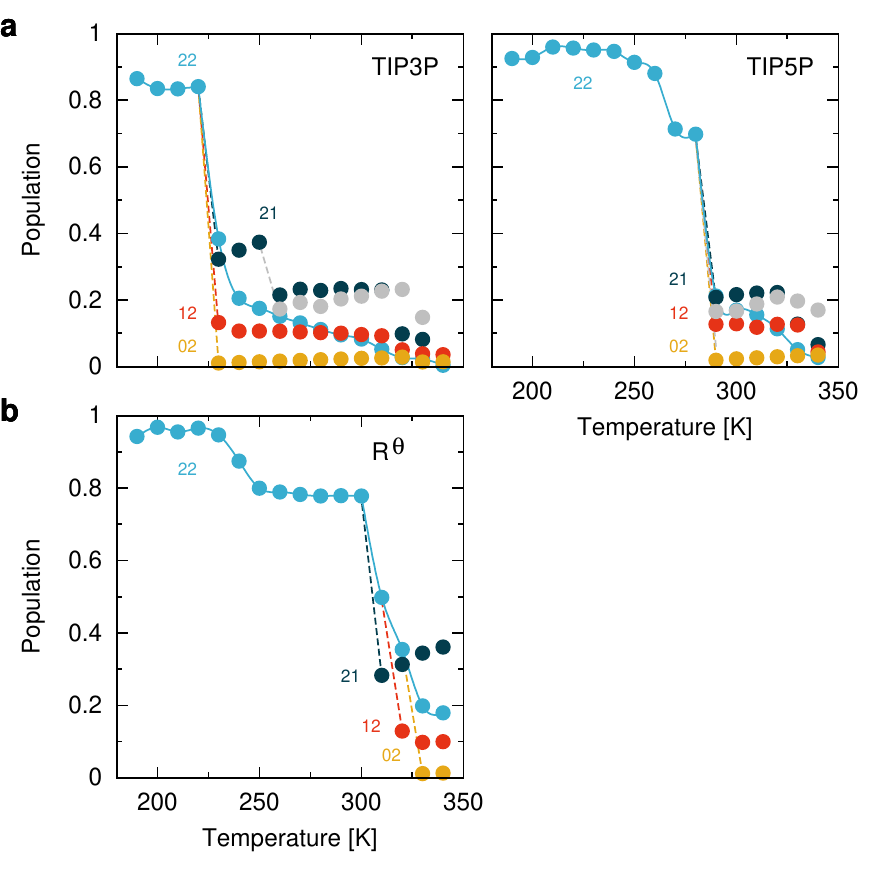}
  \caption{Robustness of the gradient-cluster analysis. (a) Gradient-cluster
  populations for TIP3P and TIP5P water models and (b) for TIP4P/2005 by using
the $R\theta$ hydrogen-bond definition.}
  \label{fig:robust}
\end{figure}

Other water models present the three free-energy landscape regimes as
TIP4P/2005. In Fig.~\ref{fig:robust}a, the population of the most visited
gradient-clusters for the TIP3P and TIP5P models are shown.
The two models are in qualitative agreement with the analysis done for
TIP4P/2005 where the exact temperature values characterizing
the three regimes depend on the specific water model used.  This is expected
due to the recently observed model dependent hydrogen-bond temperature shifts
\cite{Shevchuk2012Water}. That is, while hydrogen-bond patterns between
different water models are essentially the same, they differ by a temperature
shift of up to $\sim65$K for the TIP3P model when comparing conventional
classical water models to TIP4P/2005. For example, the appearance of the
homogeneous regime is found at 225K for TIP3P, at a roughly 60K lower
temperature than TIP4P/2005 in excellent agreement with previous analysis
\cite{Shevchuk2012Water}. Similarly, the case of TIP5P is in qualitative
agreement with TIP4P/2005.
%but a remarkable compression in temperature space (consistent with the
%increased concavity of the density function \cite{Vega2005Relation,
%Shevchuk2012Water}).

The origin of the temperature shift is related to the relative hydrogen-bond
strength differences of the various models \cite{Shevchuk2012Water}.
Consequently, it is expected that artificial modifications of the hydrogen-bond
strength due to more (or less) conservative bond definitions might shift the
three regimes as well.  This is so because water microstates are based on
hydrogen-bond connectivity and its propensity. To check this behavior, the
whole analysis was repeated by using another definition of hydrogen-bond based
on the classical inter-oxygen distance and donor-acceptor angle $R\theta$ (see
Methods).  As shown in Fig.~\ref{fig:robust}b, the overall behavior of the
gradient-cluster populations is remarkably similar with the presence of the
three liquid regimes. On the other hand, the expected temperature shift is
present.  We found that $R\theta$ predicts a larger number of hydrogen-bonds
than the Skinner definition.  For the former definition, 3.8 and
3.6 average number of hydrogen bonds per molecule are found at temperatures of
250K and 300K, respectively.  These numbers decrease to 3.7 and 3.3 when the
Skinner definition is used.  Using a less conservative definition like
$R\theta$, effectively increases hydrogen-bond strength. As a consequence, the
population of the fully-coordinated gradient-cluster is over estimated, giving
in turn a temperature shift. Being the discussion on the quality of
hydrogen-bond definitions still open \cite{Kumar2007Hydrogen,
Henchman2010Topological}, we want to remark that the change of the
hydrogen-bond definition would only slightly affect the exact position of the
three regimes but not the existence of them.

%Finally, one limitation of the present analysis is the intrinsic time
%resolution of the transition network. This is given by the time interval
%between two subsequent snapshots of the molecular dynamics trajectory. In the
%present case this interval is set to 4~fs. As such, network links bear this
%intrinsic timescale, indirectly influencing the properties of the
%gradient-clusters:  all dynamics occurring on a faster timescale would be washed
%out because barrier crossing information is lost \cite{Rao2010Protein}. 

\section{Conclusions}

We have presented extensive molecular dynamics simulations of the TIP4P/2005
model for a temperature interval ranging from 190K to 340K in conjunction with
complex network analysis to obtain a detailed mapping of the underlying
free-energy landscape. The main idea was to investigate the structural and
dynamical properties of local water configurations defined by the hydrogen-bond
connectivity with an extension of two solvation shells. 

From a microscopic point of view, the free-energy landscape of liquid water is
characterized by three major regimes.  At ambient conditions, several
metastable water configurations with distinct structure and dynamics are found
(\emph{inhomogeneous} regime). Below 285K, the free-energy landscape develops a
funnel dominated by the fully coordinated configuration with an extension of at
least two solvation shells (\emph{homogeneous} regime). By lowering the
temperature below 225K, the funnel becomes more pronounced, with the
fully-coordinated configuration becoming a global attractor of the dynamics
(homogeneous \emph{low-density} regime).

While the three regimes were deducted from water microscopic
properties, the presence of the tree regimes  is
correlated to the behavior of the density, which is an ensemble property of the
system. As such, the homogeneous low-density regime spans till the density
start to grow with a change in concavity at 225K; the homogeneous regime is
characterized by the monotonous increase of the density curve up to the density
maximum at around 280K; finally, the descending section of the density is
located into the inhomogeneous regime. 

%Another ensemble property which emerges from the microscopic analysis is the
%crossing of the temperature of maximum compressibility. This temperature is
%considered by many the signature of the Widom line, the propagation of a
%critical point at higher pressures separating two liquid phases of water.

From an experimental point of view, the presence  of structural inhomogeneities
at ambient temperature is in qualitative agreement with small-angle X-ray
scattering measurements \cite{Huang2009Inhomogeneous} while the presence of
multiple kinetics is in principle accessible to high order non-linear
spectroscopy   \cite{GarrettRoe2011ThreeDimensional}.

%Our results support the idea that free-energy representations of the
%microscopic behavior of water are a useful tool for elucidating water
%structure and dynamics at multiple levels of detail.

%The temperature of
%maximum compressibility is often associated to the Widom line of water, i.e.
%the propagation of the very much debated liquid-liquid phase transition of
%water \cite{Mishima1998Relationship, Limmer2011Putative}.

\section*{Acknowledgments}

This work is supported by the Excellence Initiative of the German Federal and
State Governments.

\bibliography{cites}

\end{document}